\documentclass[onecollarge,natbib]{svjour2}
\bibpunct{[}{]}{,}{n}{}{,} 
\smartqed  
\usepackage{graphicx}
\usepackage{mathptmx}      
\journalname{Few-Body Systems (APFB2011)}
\begin{document}

\title{\boldmath
Evidence of the non-strange partner of pentaquark
  from the elementary $K^+\Lambda$ photoproduction}

\titlerunning{Evidence for the non-strange partner of pentaquark}        

\author{T. Mart}


\institute{T. Mart \at
              Departemen Fisika, FMIPA, Universitas Indonesia,
              Depok 16424, Indonesia\\
              \email{tmart@fisika.ui.ac.id}
}

\date{Received: date / Accepted: date}

\maketitle

\begin{abstract}
 The existence of the $J^p=1/2^+$ narrow resonance predicted 
 by the chiral soliton model has been investigated by 
 utilizing the new kaon photoproduction data. For this purpose, 
 we have constructed two phenomenological models, which are 
 able to describe kaon photoproduction from threshold up to 
 $W=1730$ MeV. By varying the resonance mass, width,
 and $K\Lambda$ branching ratio in this energy range 
 we found that the most convincing mass of this resonance 
 is 1650 MeV. Using this result we estimate
 the masses of other antidecuplet family members.
\end{abstract}

\section{Introduction}
\label{intro}
The chiral soliton model proposed by Diakonov {\it et al.} 
\cite{diakonov} predicts that the non-strange partner of 
pentaquark, the $P_{11}(1710)$ narrow resonance, has
significant decay widths to the $\eta N$, $\pi N$, and $K\Lambda$
channels. The observation of pentaquark by the LEPS collaboration
almost a decade ago \cite{nakano} has also sparked considerable 
interest in the investigation of this state. It is obvious that
the $\pi N$ and $\eta N$ channels received more attention, since in both
channels there has been a large number of experimental data that has 
been precisely interpreted by a number of phenomenological models
such as MAID \cite{Drechsel:2007if} and SAID \cite{said}. 
In the $\pi N$ channel a clear 
signal of this narrow state was observed at 1680 MeV and
a weaker one was detected at 1730 MeV \cite{igor}. In the 
$\eta N$ production off a free neutron a substantial 
enhancement at $W\approx 1670$ MeV is experimentally 
found \cite{kuznetsov}. Clearly, such an enhancement 
could be explained as the presence of the narrow 
$P_{11}$ resonance. Nevertheless, a different explanation for
this enhancement is also possible, i.e., as the contributions
of the $K\Lambda$ and $K\Sigma$ loops \cite{Doring:2009qr}.

We note that there has been no attempt to study this
resonance by utilizing kaon photoproduction
prior to our previous work \cite{mart_narrow}, 
although some experimental data with relatively good quality have
been recently
provided by the CLAS \cite{CL05} and SAPHIR \cite{SP03} 
collaborations. In view of this we are interested in
following the procedure developed in 
Ref.~\cite{igor}, i.e., scanning the changes in
the total $\chi^2$ after including a $P_{11}$ narrow 
resonance with the variation of the resonance mass, width,
and $K\Lambda$ branching ratio \cite{mart_narrow}. 

Besides the difficult situation in kaon photoproduction,
the accuracy of phenomenological model plays 
a crucial role. Since the energy of interest is
very close to the $K^+\Lambda$ threshold, an accurate
model that can describe experimental data at low energies 
would be more suitable for this purpose, rather 
than a global model that fits to data in a wide energy range 
but overlooks the appearing structures at low energies. 

In this paper we present 
further results of our calculation which support the evidence
of this narrow state. Using this result we estimate the masses
of other antidecuplet family members by utilizing the mass splitting 
of 110 MeV, which is theoretically predicted in Ref.~\cite{igor}.

\section{The isobar Model}
\label{sec:model}
For the purpose of investigating kaon photo- and electroproduction
near threshold,  in the previous work 
\cite{Mart:2010ch} we have constructed an isobar model from 
the standard $s$-, $u$-, and $t$-channel Born terms along 
with the $K^{*+}(892)$ and $K_1(1270)$ $t$-channel vector 
mesons, as well as an  $S_{11}(1650)$ nucleon- and 
an $S_{01}(1800)$ hyperon-resonance. The latter
is included in order to improve the agreement with experimental 
data. The model fits nicely all available data from threshold 
($W=1609$ MeV) up to
$W=1660$ MeV. However, as predicted by many soliton models 
the most convincing mass of the narrow resonance
is around 1680 MeV \cite{igor,Walliser:2003dy}. Therefore, an extension
of the model to cover energies between threshold and $W=1730$ 
MeV is mandatory. Fortunately, at this energy regime 
both experimental data from SAPHIR and CLAS collaborations
are in agreement with each other and, consequently, 
the problem of data inconsistency investigated in
Ref.~\cite{Mart:2006dk} does not exist. Moreover, the 
hadronic form factors required to suppress the diverging 
Born terms would play a less significant role here.

\begin{figure*}
\centering
  \includegraphics[width=1.0\textwidth]{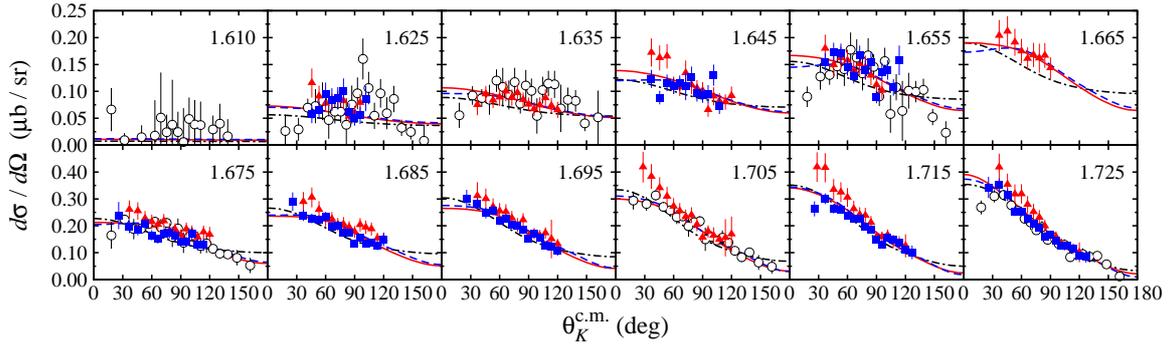}
\caption{(Color online) Comparison between angular distributions of 
  differential cross section obtained from Model 1
  (solid lines), Model 2 (dashed lines), and Kaon-Maid 
  (dash-dotted lines) \cite{kaon-maid} with experimental data.
  Notation of experimental data can be found
  in Ref.~\cite{Mart:2010ch}.
  The corresponding total c.m. energy $W$ (in GeV) is
  shown in each panel.}
\label{fig:dkpl}
\end{figure*}

\begin{figure*}[b]
\centering
  \includegraphics[width=0.85\textwidth]{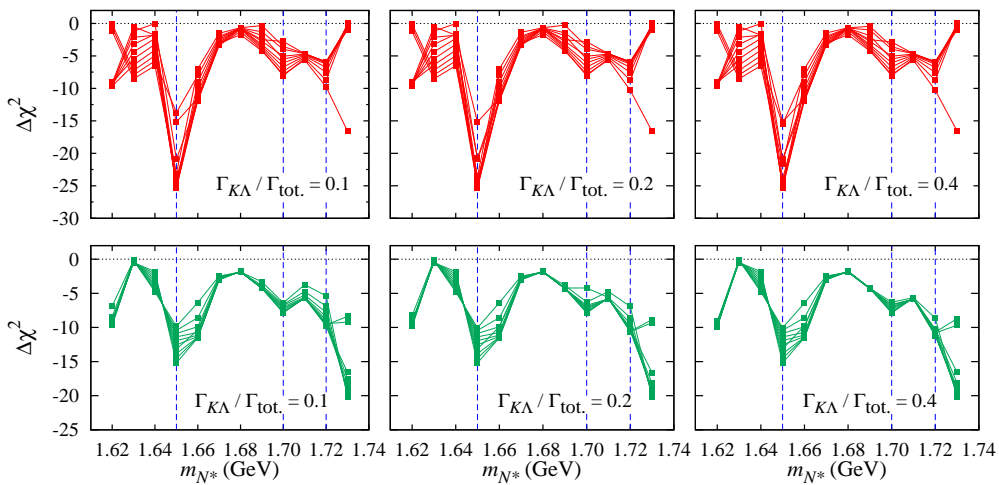}
\caption{(Color online) Change of the $\chi^2$ in the fit of Model 1 (top panels) 
	and Model 2 (bottom panels) due to the inclusion of the
	$P_{11}$ resonance with the mass scanned from 1620 
	to 1730 MeV (step 10 MeV) and $\Gamma_{\rm tot.}$ taken from
	1 to 10 MeV (step 1 MeV) for different $K\Lambda$ branching
	ratios ($\Gamma_{K\Lambda}/\Gamma_{\rm tot.}= 0.1$, 
        0.2, and 0.4). The three vertical lines in each panel 
        indicate the values of $m_{N^*}=1650$,
	1700 and 1720 MeV.}
\label{fig:scan}
\end{figure*}

In the energy range of interest there exist six nucleon 
resonances which may contribute to this process.
Their properties 
relevant to the present work are mostly available from the 
Particle Data Book (PDG) \cite{nakamura}.
Other unknown coupling constants can be fitted from experimental
data. The result of the fit is shown in Fig.~\ref{fig:dkpl},
where the prediction of Kaon-Maid \cite{kaon-maid} is also
shown for comparison. To investigate the model dependence
of the result in the next Section, here we propose two 
models. In Model 1 we restrict the maximum variation of 
the photon amplitudes during the fitting process to 10\% 
of the original PDG values, whereas in Model 2 
all parameters are allowed to vary within the PDG error
bars. From Fig.~\ref{fig:dkpl} it is clear that both 
models display a good agreement with experimental data
and might provide a significant improvement of Kaon-Maid in the
energy range of interest. Results of both models for other
polarization 
observables can be found in our previous paper \cite{mart_narrow}.

\section{Searching for the narrow resonance}
In order to observe the existence of a $P_{11}$ narrow resonances
in kaon photoproduction we scan the changes in 
the total $\chi^2$ after including this  
resonance with the variation of its mass, 
width (1 to 10 MeV with 1 MeV step),
and $K\Lambda$ branching ratio.
The results for both Model 1 and 2 are shown in 
Fig.~\ref{fig:scan}.
In all three values of the $K\Lambda$ branching
ratios selected, 
we can see that three minima at $m_{N^*}=1650$, 1700, and 1720 MeV
appear consistently. Nevertheless, the minimum $\Delta\chi^2$ at
$m_{N^*}=1650$ MeV seems to be the most convincing one. 
It is found that the lowest values of
$\Delta\chi^2$ can be obtained by using $\Gamma_{\rm tot.}=5$ MeV.
Variation of the $K\Lambda$ branching ratio changes these 
values only slightly. We have also investigated the possibility
that the extracted resonance not a $P_{11}$ state, but an $S_{11}$
or even a $P_{13}$ state. It is shown that the latter is less likely,
whereas most available observables are able to distinguish the effect 
of $S_{11}$ and $P_{11}$ states \cite{mart_narrow}. 

\begin{figure*}
\centering
  \includegraphics[width=0.82\textwidth]{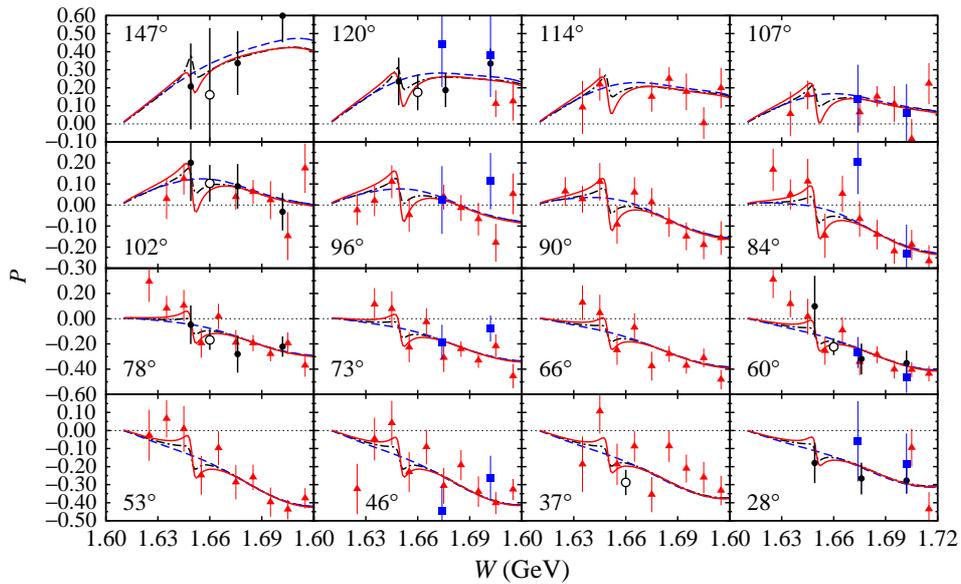}
\caption{(Color online) Effects of the inclusion of $P_{11}$ (solid lines) 
  and $S_{11}$ (dash-dotted lines) states on the energy distribution of 
  the $\Lambda$ recoil polarization investigated by using Model 1
  for different kaon scattering angles (shown in each panel).
  The dashed lines show the result without narrow resonance
  in the model. 
  Notation of experimental data can be found
  in Ref.~\cite{Mart:2010ch}.}
\label{fig:polar}
\end{figure*}

To investigate model dependence of our result we display the
the same changes in the total $\chi^2$, 
but calculated by using Model 2,  in the 
bottom panels of Fig.~\ref{fig:scan}. Once again, we see a similar
pattern as in Model 1. We, therefore,
conclude that the minimum at $m_{N^*}=1650$ seems to be 
model independent, whereas the minima at $1700$ and 1720 MeV 
become much weaker in Model 2. Although this might imply that 
the possibility of a narrow $P_{11}$ resonance with a mass 
of 1700 or 1720 MeV could not be excluded, 
we believe that investigation of this resonance 
at energies around 1700 MeV by using the present mechanism is 
difficult due to the opening of 
$K\Sigma$, $\rho p$, and $\omega p$  channels.
It is also important to mention
here that by including the $P_{11}$ state in Model 1 
the number of fitted parameters increases from 41 to 45, whereas the total
$\chi^2$ decreases from 859 to 834. This corresponds to
a statistical significance\footnote{The author 
is indebted to Prof. Takashi Nakano 
for clarifying this issue.} of $4\sigma$.

\begin{figure*}
\centering
  \includegraphics[width=0.45\textwidth]{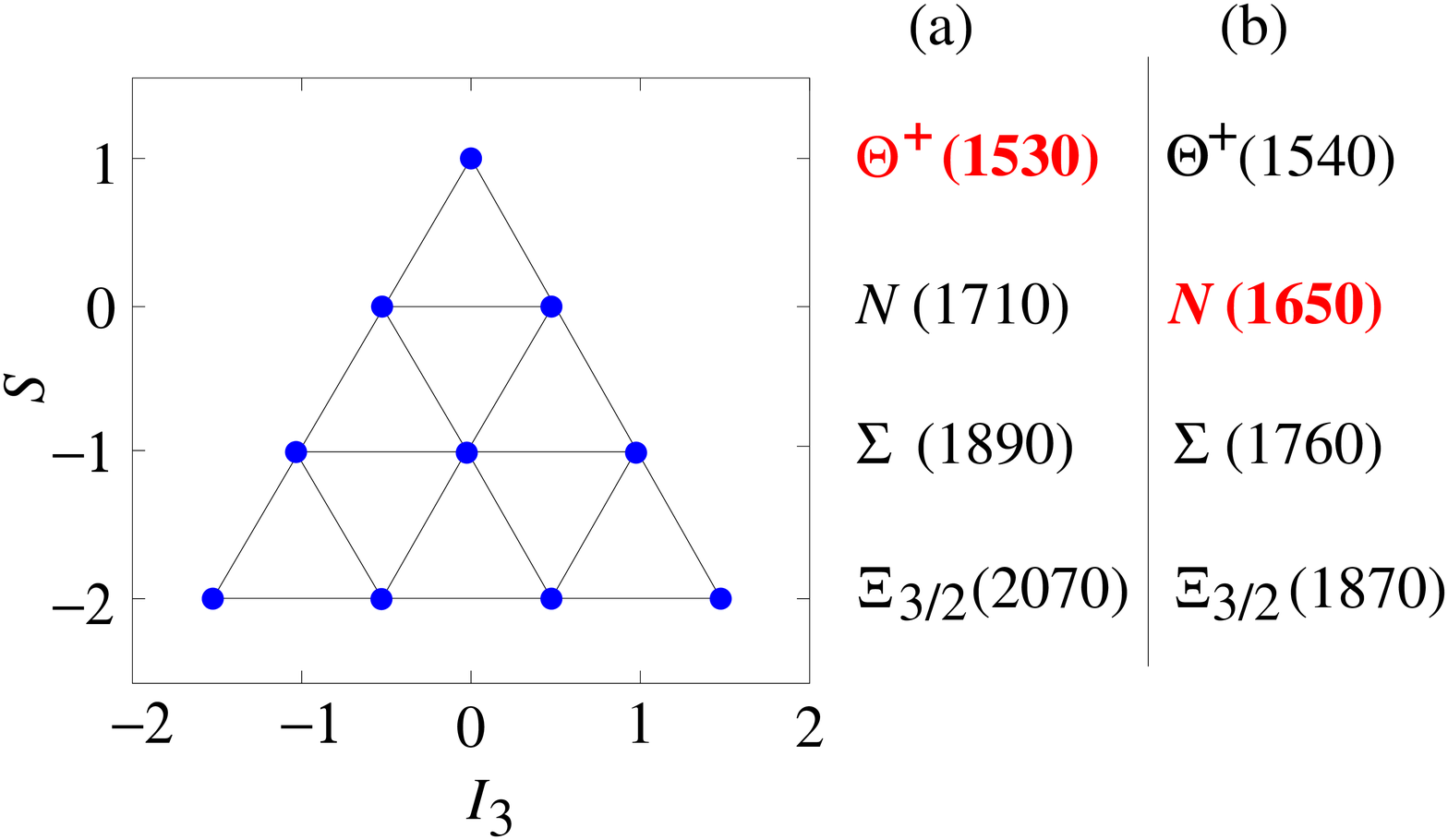}
\caption{(Color online) The masses of the antidecuplet member as suggested by
 (a) Ref.~\cite{diakonov} and (b) the present work with mass
 splitting of 110 MeV. Boldface fonts indicate inputs of 
 the calculations.}
\label{fig:antidecuplet}
\end{figure*}

By scrutinizing the contributions of individual data 
to the $\chi^2$ in our fits we found that the minimum 
at 1650 MeV originates mostly from the $\Lambda$ recoil 
polarization data \cite{McCracken:2009ra}
as shown in Fig.~\ref{fig:polar}.
From this figure we can see that there exists a dip at $W\approx 1650$ MeV
in the whole angular distribution of data. It is also
apparent that both $P_{11}$ and $S_{11}$ states can nicely describe 
this dip. Although the recoil polarization is 
not a suitable observable to distinguish different states at
1650 MeV, more precise data in this case 
are still urgently required 
to remove uncertainties in the position of the dip.

\section{Further consequence}
Although previous investigations using pion and eta 
photoproductions obtained the  
$P_{11}$ mass around 1680 MeV, the 1650 MeV 
mass obtained in the present work corroborates the  
calculation utilizing the Gell-Mann-Okubo 
rule \cite{igor} and is in a good agreement with the 
prediction of the topological soliton model of 
Walliser and Kopeliovich~\cite{Walliser:2003dy}.
Nevertheless, all previous calculations have used the
pentaquark mass obtained by Nakano {\it et al.} \cite{nakano}
in order to determine the masses of the whole antidecuplet
family members. It is naturally interesting to ask: how this 
result would change if we used the mass of the non-strange
member $P_{11}$ obtained in the present work to estimate
the masses of the rest family members. The answer is given
in Fig.~\ref{fig:antidecuplet}, where we compare the masses
predicted by Diakonov {\it et al.} \cite{diakonov} with those of
the present result. Note that in obtaining
this result we have used the mass splitting proposed in 
Ref.~\cite{igor} (i.e. 110 MeV), which is also
compatible with the prediction of Walliser and Kopeliovich
\cite{Walliser:2003dy}. As expected, the mass of the pentaquark
does not change from its original value \cite{nakano}, whereas
the masses of other family members are significantly reduced.
However, it is important to emphasize here that 
our finding is in a good agreement
with the theoretical prediction of Ref. \cite{igor}.

\begin{acknowledgements}
Supports from the University of Indonesia
and the Competence Grant of the Indonesian 
Ministry of National Education are gratefully
acknowledged.

\end{acknowledgements}

\end{document}